# Excitation Spectrum and Stability from a Filled Landau Level in Rotating Dipolar Fermi Gases

Szu-Cheng Cheng

Department of Physics, Chinese Culture University, Taipei, Taiwan, R. O. C.

We apply the equation-of-motion method to study the collective excitation spectrum from a filled Landau level in rotating dipolar Fermi gases. The predicted excitation spectrum of rotating dipolar Fermi gases can exhibit a roton-minimum character. This roton character is tunable by varying the dipole interaction strength and confining potential. An increase of the dipole interaction strength makes the roton minimum becoming zero, and the system becomes unstable. We also obtain a condition for the dynamical stability of rotating dipolar Fermi gases.

Recently there is a remarkable progress in the experiment to manipulate quantum many-body states in the ultra cold atomic gases. Systems with well understood and controllable interactions are available. The richer and more complex many-body states in the strong interaction regime are observable from the achievement involving Feshbach resonances [1] and the realization of a Mott insulator-superfluid transition of atoms in optical lattices [2]. It is also possible to realize strongly correlated states of the fractional quantum Hall effect (FQHE) in a rotating Bose gas [3]. Theoretically, we expect the existence of the fractional-quantum-Hall states in the context of ultra cold atomic gases with short-range contact interactions [3]. In fact, it is difficult to realize the FQHE states experimentally, due to a small excitation gap with weak contact interactions. Since interactions between atoms localized in the lattices being strongly enhanced compared to the interaction of atoms in a trap, a novel method that uses atoms in optical lattice to raise the excitation gap of the system is proposed [4].

It is also possible to explore the potentially strong correlations induced by the long-range interaction. The recent realization of a concrete physical system plays a significant role in the creation of the dipolar interaction [5]. The dipolar interaction is a long-range interaction that puts dipolar gases correlated strongly [6]. There is the experimental feasibility of incompressible fractional quantum Hall-like state in ultracold two-dimensional rapidly rotating dipolar Fermi gases [7]. The intra-Landau-level excitation spectrum of rotating dipolar Fermi gases exhibited a finite energy gap in the long wavelength limit and a roton minimum at finite wave vector [8]. Rotating gases feel the Coriolis force in the rotating frame. The Coriolis force on rotating gases is identical to the Lorentz force of a charged particle in a magnetic field. Quantum-mechanically, energy levels of a charged particle in a uniform magnetic field are named as Landau levels and display discrete energy levels of a harmonic oscillator. There is a large degeneracy in the Landau level. It is convenient to define the filling factor $\nu$, $\nu \equiv 2\pi\rho a^2$, to measure the fraction of Landau levels being occupied by particles, where $\rho$ and $a$ stand the average density and the magnetic length, respectively. The discreteness of the eigenvalue spectrum is essential for the occurrence of the integral quantum Hall effect [9]. Filling factor ν is an integral number as Landau levels are completely filled by fermions in the integral quantum Hall effect. In the filled-Landau-level case, due to the Pauli Exclusion Principle, the only excitations are inter-Landau-level excitations involving the promotion of a particle from an occupied Landau level to an unoccupied Landau level [10, 11].

In this letter we report the first evaluation of inter-Landau-level excitation energies of the filled Landau states in quasi two-dimensional (2D) dipolar gases whose mass and dipole moment are $M$ and $d$, respectively. Dipolar gases are rotating and fully polarized along the rotating axial direction. The approach is similar to the equation-of-motion (EOM) method used in calculations of inter-Landau-level excitation energies of 2D interacting systems [12]. We focus on low-energy excitations of the density oscillations in the limit where the ground state is the state with Landau levels filled. To avoid the inhomogeneous effect in the 2D direction, we studied the system rotating in the limit of critical rotation, where the magnitude of the rotating frequency $\omega$ of the system is close to but still smaller than the trapping frequency. Under the limit of critical rotation, the density of the trapped gas becomes uniform except at the boundary given by a trapping potential.

There are many different ways to calculate the excitation spectrum of any system, the Green's function method [10], the time-dependent Hartree-Fock theory [11], and the method of linearized EOM [13]. The time-dependent Hartree-Fock theory appears to be a natural extension of the stationary Hartree-Fock picture; it allows a self-consistent oscillation of the potential with the particles. It is equivalent to the linearized EOM method. The linearized EOM method involves a rather arbitrary linearization procedure which tends to be a prescription rather than a derivation. In this paper we are going to use the EOM method to calculate the excitation spectrum from a filled Landau level of rotating dipolar Fermi gases. The EOM method [14] described in this paper is a particular simple one which shows much greater flexibility than the linearized EOM method. It differs from the linearized EOM method in certain fundamental respects. The EOM are not expressed as operator equations, but rather as the ground-state expectation of operator equations. Thus all arbitrariness of linearization procedures is avoided.

The formal development of the EOM method for a 2D electron gas at strong magnetic field has been presented elsewhere [12]. We shall apply previous formalisms and take the cyclotron energy $\hbar\omega_c = 1$ (or $h\nu_c = 1$) and the magnetic length $a = \sqrt{\hbar/M\omega_c} = 1$, where the cyclotron frequency $\omega_c = 2\omega$ and $\nu_c = \omega_c/2\pi$. The Hamiltonian of a quasi-2D rotating dipolar Fermi gas is

$$H = \frac{1}{2}\sum_\mu \left[-i\nabla_\mu + \mathbf{A}(\mathbf{r}_\mu)\right]^2$$

$$+ \frac{1}{2}\Omega L^{-2} \sum_{\mathbf{q}} \sum_{\mu \neq \lambda} V(\mathbf{q})\exp[i\mathbf{q}\bullet(\mathbf{r}_\mu - \mathbf{r}_\lambda)] , \qquad (1)$$

where $\mu$ and $\lambda$ are the index of particles, $\mathbf{q}$ is the wave vector, $L$ is the length of the system and $\Omega \equiv (d^2/a^3)/\hbar\omega_c$. We chose the Landau gauge for the vector potential $\mathbf{A}(\mathbf{r}) = -y\hat{x}$ along the $x$-direction. The $V(\mathbf{q})$ in Eq. (1) is the Fourier transform of the quasi-2D dipolar-interaction potential [15]: $V(\mathbf{q}) = -2\pi q \exp(\xi^2) \text{Efrc}(\xi)$, where $\xi = qz/\sqrt{2}a$, $z$ is the extension of dipolar gases in the axial direction and $\text{Erfc}(\xi) = 1 - \left(2/\sqrt{\pi}\right)\int_0^\xi \exp(-t^2)dt$. Note that the contact interaction term in $V(\mathbf{q})$ is ignored due to the Pauli Exclusion Principle of fermions.

Let $O^\dagger$ be the excitation operator. The $O^\dagger$ operator has to satisfy the pseudo-boson commutator: $\langle\varphi|[O, O^\dagger]|\varphi\rangle = 1$, where $|\varphi\rangle$ is the ground-state wave function. The excitation energy $\Delta$ of a single mode is given by

$$\langle\varphi|[O, H, O^{\dagger}]|\varphi\rangle = \Delta\langle\varphi|[O, O^{\dagger}]|\varphi\rangle ,  \quad (2)$$

and the double commutator $[A, B, C]$ is defined as $2[A, B, C]=[A, [B, C]]+[[A, B], C]$. Equation (2) is just the EOM proposed by Rowe [14] and has been used extensively in nuclear and molecular physics. In Eq. (2) one must specify which type of excitation is included in the excitation operator $O^{\dagger}$ and what approximate ground-state wave function $|\varphi\rangle$ will be used to evaluate the expectation values of the commutators. For our case we can approximate $|\varphi\rangle$ by a dipolar Fermi gas rotating with a frequency such that Landau levels are completely filled. In the fractional-quantum-Hall effect, an excited state constructed from the density operator, which is in spirit similar to Feynman's $^4$He theory, is used to study the collective-excitation spectrum of a 2D electron gas in a magnetic field [16] and a 2D rotating dipolar Fermi gas [8]. We also construct the excitation operator $O^{\dagger}(\mathbf{q})$ at wave vector $\mathbf{q}$ from the Fourier transform of the density operator $\rho_{\mathbf{q}} = \sum_{\mu} \exp[i\mathbf{q}\cdot\mathbf{r}_{\mu}]$. Since the restriction to an integral number of the filled Landau levels is made, there are no intra-Landau-level excitations. The collective excitations of the system are the familiar excitation modes in which one particle is excited to an unoccupied Landau level $m$, leaving behind a hole in a filled Landau level $j$. Using the Landau wave function as a basis, the excitation operator $O^{\dagger}(\mathbf{q})$ in the Landau gauge is given by $O^{\dagger}(\mathbf{q}) = \sum_{m,j}[X_{mj}(q)B^{\dagger}_{mj}(\mathbf{q}) - Y_{mj}(q)B_{mj}(-\mathbf{q})]$ and

$$B^{\dagger}_{mj}(q) = \sqrt{2\pi/L^2} \sum_{p,k} \delta(p-k+q_x)$$

$$\times \exp[ipq_y - i(m-j)\theta + iq_x q_y/2]\, c^{\dagger}_{mk} c_{jp}, \quad (3)$$

where $p$ and $k$ are the wave vectors in $x$-direction and $\theta = \tan^{-1}(q_y/q_x)$. $c^{\dagger}$ and $c$ are the creation and annihilation operator of particles, respectively. Here (and throughout the rest of the paper) the subscripts $m$ and $n$ are reserved for the Landau levels above the Fermi energy, and $\ell$ and $j$ for the Landau levels below the Fermi energy.

Operator $O^\dagger(\mathbf{q})$ has two separate terms: $X_{mj}(q)$ terms create particle-hole states and $Y_{mj}(q)$ terms destroy particle-hole states to show the correlation of the system. From the excitation operator $O^\dagger(\mathbf{q})$ satisfying the pseudo boson commutator, we have $\sum_{m,j} [X_{mj}(q)X_{mj}^*(q') - Y_{mj}(q)Y_{mj}^*(q')] = \delta_{\mathbf{qq'}}$.

Inserting $O^\dagger(\mathbf{q})$ into Eq. (1), we obtain a matrix equation

$$\begin{bmatrix} D & F \\ F^\dagger & D^* \end{bmatrix} \begin{bmatrix} X \\ Y \end{bmatrix} = \Delta \begin{bmatrix} X \\ -Y \end{bmatrix}, \qquad (4)$$

where $X$ and $Y$ are the column vectors $[X_{mj}(q)]$ and $[Y_{mj}(q)]$, respectively. The matrix $D$ is Hermitian and defined by $D_{mj,n\ell}(\mathbf{q}) = \langle \varphi | [B_{mj}(\mathbf{q}), H, B_{n\ell}^\dagger(\mathbf{q})] | \varphi \rangle$; the matrix $F$ is a symmetric and defined by $F_{mj,n\ell}(\mathbf{q}) = -\langle \varphi | [B_{mj}(\mathbf{q}), H, B_{n\ell}(-\mathbf{q})] | \varphi \rangle$. Eq. (4) is not a Hermitian matrix. The usual way to diagonalize a Hermitian matrix is not applicable in diagonalizing Eq. (4). A simple diagonalization procedure for Eq. (4) is derived thoroughly by Ullah and Rowe [17].

Using the method from Ullah and Rowe [17], numerical solutions of Eq. (4) for $\nu=1$ and $\nu=2$ cases are shown in figures 1 and 2, respectively. We studied the collective-mode dispersion for different values of the dimensionless thickness parameter $z/a$ along the rotating axis. The essential features of excitation dispersions exhibit a finite excitation-energy gap $h\nu_c$ at $q=0$ and energy minima, called roton modes, at finite wave vector. The lowest excitation mode at $q=0$ is always $h\nu_c$, as a consequence of Kohn's theorem [18]. This result indicates that our calculation is checked by Kohn's theorem. In Fig. 1, one roton mode that is close to $qa=2$ is observed at $\nu=1$. For $\nu=2$, the excitation spectrum shows multi roton minima in Fig. 2. These roton modes become soft and lie below $h\nu_c$ as the dipole interaction strength is increasing. The depth of roton minima becomes shallower as $z/a$ is larger. Therefore, the correlation effect induced by the dipole interaction is decreasing while $z/a$ is increasing. An increase of the dipole interaction strength makes the roton minimum deeper. For $\nu=1$, $\Omega=1.48$ and $z/a=0$ ($\nu=2$, $\Omega=1.92$ and $z/a=0$) the minimum energy reaches zero at $qa\sim2$ ($qa\sim1.4$). At larger values of $\Omega$ one gets imaginary excitation energies of $\nu=1$ and $\nu=2$ for $qa\sim2$ and $qa\sim1.4$, respectively, and the system becomes unstable. The critical values of occurring instability have been calculated numerically as functions of $z/a$, and are shown in Fig. 3.

The instability with regard to short-wave excitations from a filled Landau level in rotating dipolar Fermi gases is related to the momentum dependence of the dipole interaction strength. The roton minimum reaches zero at a given $\Omega$ just below the point of instability indicates that there is a new ground state in the region of the instability of rotating dipolar Fermi gases. The new ground states related to $\nu=1$ and $\nu=2$ are different. We interpret the collapse of the roton minimum at a given $\Omega$ just below the point of instability to be a precursor of forming a Wigner crystal [16]. The evidence in favor of this interpretation is provided by the fact that the unstable excitation is the one with momentum lying close to the primitive reciprocal-lattice wave vector $G_1$ of corresponding Wigner crystal, where $G_1 = 2.69qa$. For $\nu=2$ the momentum of the unstable mode is lying far away from the primitive reciprocal-lattice wave vector $G_2 = 3.81qa$ of corresponding Wigner crystal. Therefore, the new ground state for $\nu=2$ is not a Wigner crystal. It is likely that the ground state in higher Landau levels is a bubble crystal [19]. If there are 7 particles per bubble to form a bubble crystal of triangular structure for $\nu=2$, the primitive reciprocal-lattice wave vector of corresponding bubble crystal will be $1.44qa$, which is lying close to the momentum of the unstable mode (see Fig. 2). Although we have identified the new ground state for $\nu=1$ is a Wigner crystal, we can not rule out the possible existence of a bubble phase. If there are 2 particles per bubble to from a bubble crystal for $\nu=1$, the primitive reciprocal-lattice wave vector of corresponding bubble crystal will be $1.90qa$, which is also lying close to the momentum of the unstable mode (see Fig. 1). The presence and its properties of a bubble phase or a Wigner crystal will be a subject of our future studies.

Having understood a physical picture of excitations, we now discuss the possible experimental ways to verify theoretical results. The presence of maxima and minima in the excitation spectrum can be observed from Bragg spectroscopy [20]. Bragg spectroscopy of a Bose-Einstein condensate in a trap has been realized experimentally [21]. The measured excitation spectrum agreed well with the Bogoliubov spectrum for a condensate. Hafezi *et al*. [22] also proposed a detection technique based on Bragg spectroscopy to obtain the excitation spectrum and the static structure factor of the FQHE in an optical lattice. Their study has shown that excitation gaps of the FQHE can be realized by Bragg spectroscopy.

In conclusion, we investigated the excitation spectrum from a filled Landau level in rotating dipolar Fermi gases. The finite-thickness effects on excitations were considered. We have found that rotating dipolar Fermi gases can exhibit a gap $h\nu_c$ at the long wavelength limit and a roton character of the excitation spectrum. The depth of the roton minimum is tunable by varying the dipole interaction strength and confining potential. An increase of the dipole interaction strength makes the roton minimum deeper and lying below $h\nu_c$. At critical interaction strength the roton minimum reaches zero, and the system becomes unstable. We presented a condition for the dynamical stability of rotating dipolar Fermi gases.

The author acknowledges the financial support from the National Science Council (NSC) of Republic of China under Contract No. NSC96-2112-M-034-002-MY3. The author also thanks the support of the National Center for Theoretical Sciences of Taiwan during visiting the center.

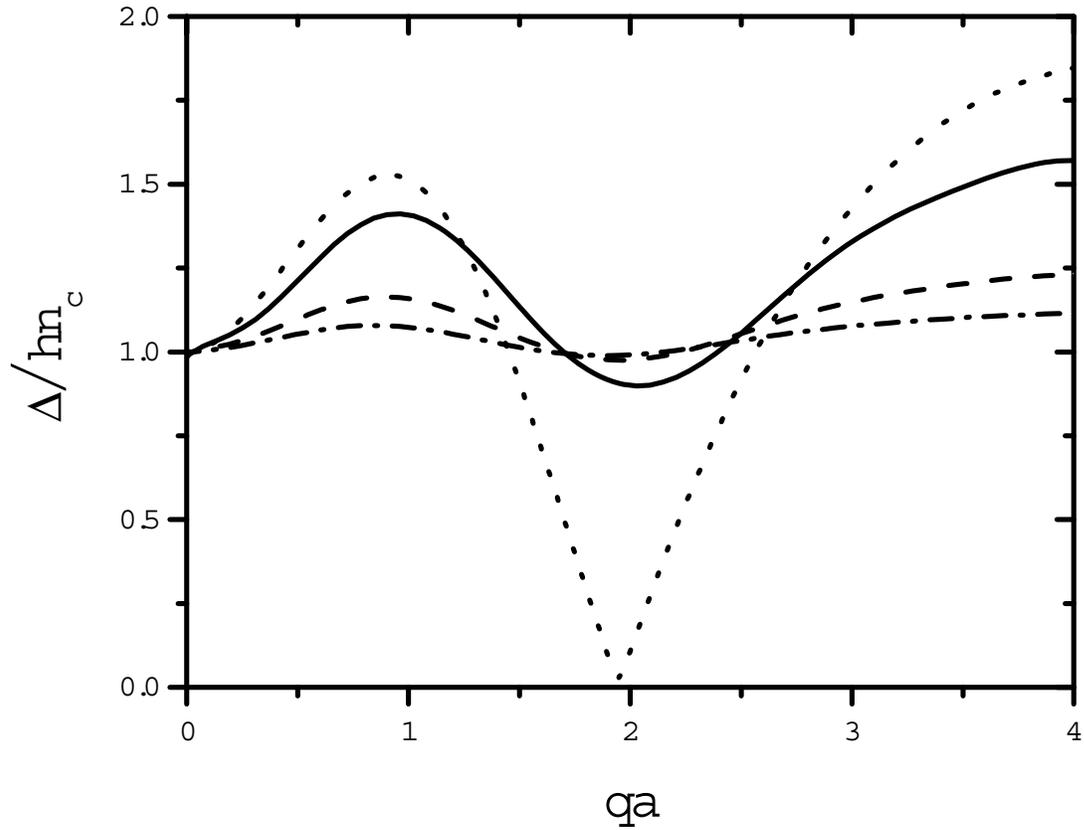

Fig. 1. Collective-excitation dispersions at $\nu=1$. Going from the top curve to the bottom, the parameters of dispersions are $\Omega=1.48$, $z/a=0.0$ (dotted line), $\Omega=1.00$, $z/a=0.0$ (solid line), $\Omega=1.00$, $z/a=0.4$ (dashed line), $\Omega=1.00$, $z/a=0.8$ (dash dotted line), respectively. The magnitude of primitive reciprocal-lattice wave vector of corresponding Wigner crystal is at $2.69qa$.

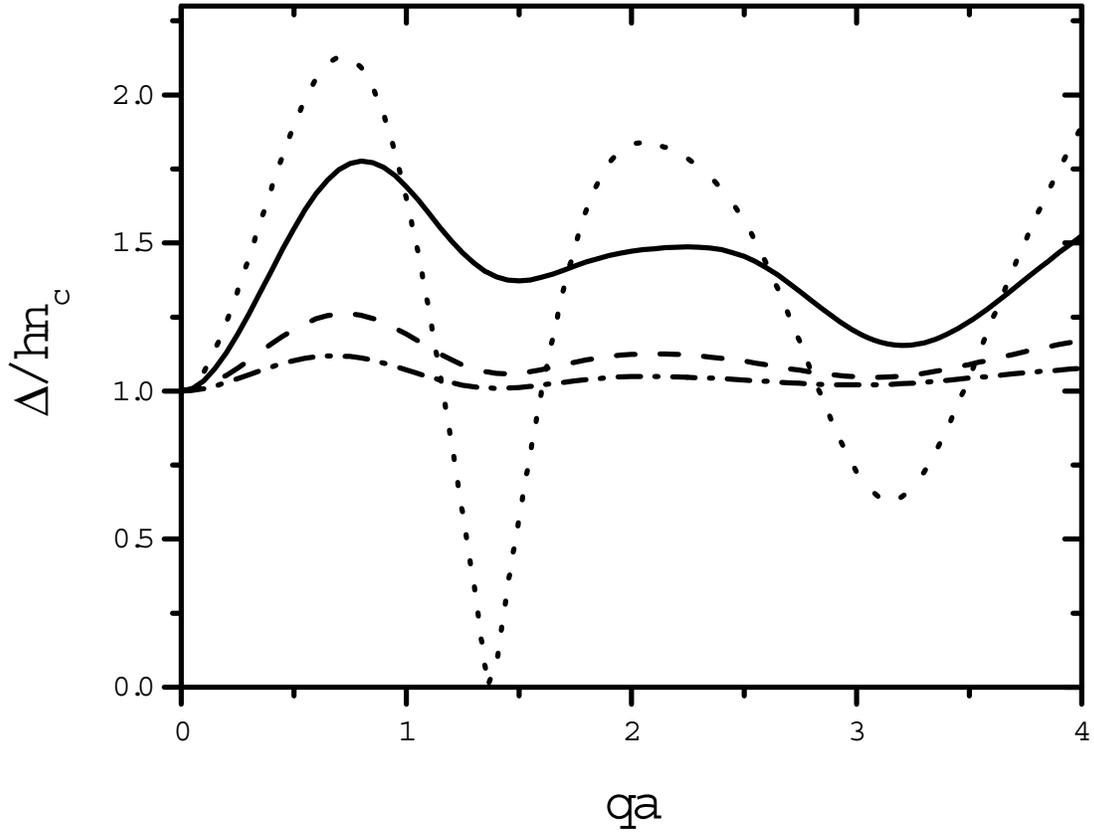

Fig. 2. Collective-excitation dispersions at $\nu=2$. Going from the top curve to the bottom, the parameters of dispersions are $\Omega=1.92$, $z/a=0.0$ (dotted line), $\Omega=1.00$, $z/a=0.0$ (solid line), $\Omega=1.00$, $z/a=0.4$ (dashed line), $\Omega=1.00$, $z/a=0.8$ (dash dotted line), respectively. The magnitude of primitive reciprocal-lattice wave vector of corresponding bubble crystal is at $1.44qa$.

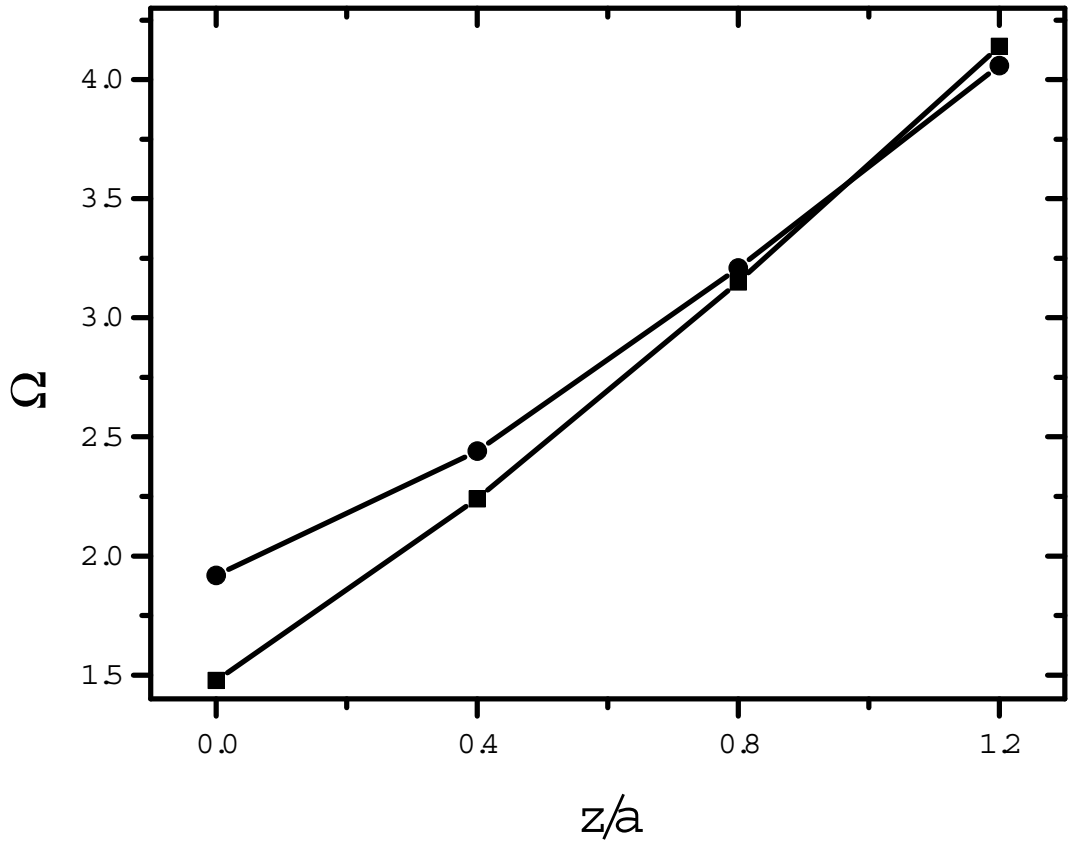

Fig. 3. Critical interaction strength for dynamical instability. Squares and circles are the values of the critical interaction strength at $\nu=1$ and $\nu=2$, respectively.

13